\documentclass[12pt,preprint,super,numbers]{aastex}

\def\msol{\,\mathrm{M}_\odot}

\newcommand{\gaia}{{\it Gaia~}}

\renewcommand{\araa}{Ann. Rev. Astron.~Astrophys.}
\renewcommand{\aap}{Astron.~Astrophys.}
\renewcommand{\aapr}{Astron.~Astrophys. Rev.}
\renewcommand{\apj}{Astrophys. J.}
\renewcommand{\apjl}{Astrophys. J. Lett.}
\renewcommand{\apjs}{Astrophys. J. Suppl.}
\renewcommand{\aj}{Astron. J.}
\renewcommand{\mnras}{Mon.~Not.~R.~Astron.~Soc.}

\usepackage{amsmath}
\usepackage{textcomp}
\usepackage{amssymb}
 \usepackage{multirow}
\usepackage{booktabs}
\usepackage{graphicx}
\usepackage[export]{adjustbox}
\usepackage[superscript,biblabel]{cite}

\newdimen\minuswidth    
\setbox0=\hbox{$-$}
\minuswidth=\wd0
\catcode`@=\active
\def@{\kern\minuswidth}
\setbox0=\hbox{\rm0}

\shorttitle{}
\shortauthors{Helmi et al.}

\begin{document}
\title{The merger that led to the formation of the Milky Way's inner
  stellar halo and thick disk}

\author{{\rm
Amina Helmi\altaffilmark{1}, 
Carine Babusiaux\altaffilmark{2}, 
Helmer H. Koppelman\altaffilmark{1}, Davide Massari\altaffilmark{1},  Jovan Veljanoski\altaffilmark{1}, 
Anthony G. A. Brown\altaffilmark{3}
}}
\affil{\altaffilmark{1}{\rm \it \footnotesize Kapteyn Astronomical Institute, University of Groningen,
P.O. Box 800, 9700 AV Groningen, The Netherlands}}
\affil{\altaffilmark{2}{\rm \it \footnotesize Univ. Grenoble Alpes, CNRS, IPAG, 38000 Grenoble, France and GEPI, Observatoire de Paris, Universit\'{e} PSL, CNRS, 5 Place Jules Janssen, 92190 Meudon, France}}
\affil{\altaffilmark{3}{\rm \it  \footnotesize Leiden Observatory, Leiden University, P.O. Box 9513, 2300 RA Leiden, The Netherlands}}


{\bf The assembly process of our Galaxy can be retrieved using the
  motions and chemistry of individual
  stars\cite{freeman-BH,hws2003}. Chemo-dynamical studies of the
  nearby halo have long hinted at the presence of multiple components
  such as streams\cite{h99}, clumps\cite{morrison2009},
  duality\cite{carollo2007} and correlations between the stars'
  chemical abundances and orbital
  parameters\cite{Chiba-Beers2000,nissen2010,Beers2017}. More
  recently, the analysis of two large stellar
  surveys\cite{apogee-dr14,prusti16} have revealed the presence of a
  well-populated chemical elemental abundance
  sequence\cite{nissen2010,hayes2018}, of two distinct sequences in
  the colour-magnitude diagram\cite{Babusiaux2018}, and of a prominent
  slightly retrograde kinematic
  structure\cite{belokurov2018,helmer2018} all in the nearby halo,
  which may trace an important accretion event experienced by the
  Galaxy\cite{haywood2018}. Here report an analysis of the kinematics,
  chemistry, age and spatial distribution of stars in a relatively
  large volume around the Sun that are mainly linked to two major
  Galactic components, the thick disk and the stellar halo. We
  demonstrate that the inner halo is dominated by debris from an
  object which at infall was slightly more massive than the Small
  Magellanic Cloud, and which we refer to as Gaia-Enceladus.  The
  stars originating in Gaia-Enceladus cover nearly the full sky, their
  motions reveal the presence of streams and slightly retrograde and
  elongated trajectories.  Hundreds of RR Lyrae stars and thirteen
  globular clusters following a consistent age-metallicity relation
  can be associated to Gaia-Enceladus on the basis of their
  orbits. With an estimated 4:1 mass-ratio, the merger
  with Gaia-Enceladus must have led to the dynamical heating of the
  precursor of the Galactic thick disk and therefore contributed to the
  formation of this component approximately 10 Gyr ago. These findings
  are in line with simulations of galaxy formation, which
  predict that the inner stellar halo should be dominated by debris
  from just a few massive progenitors\cite{hws2003,cooper2010}.}

\begin{figure}[!h!t]
\centering%
\includegraphics[scale=0.3]{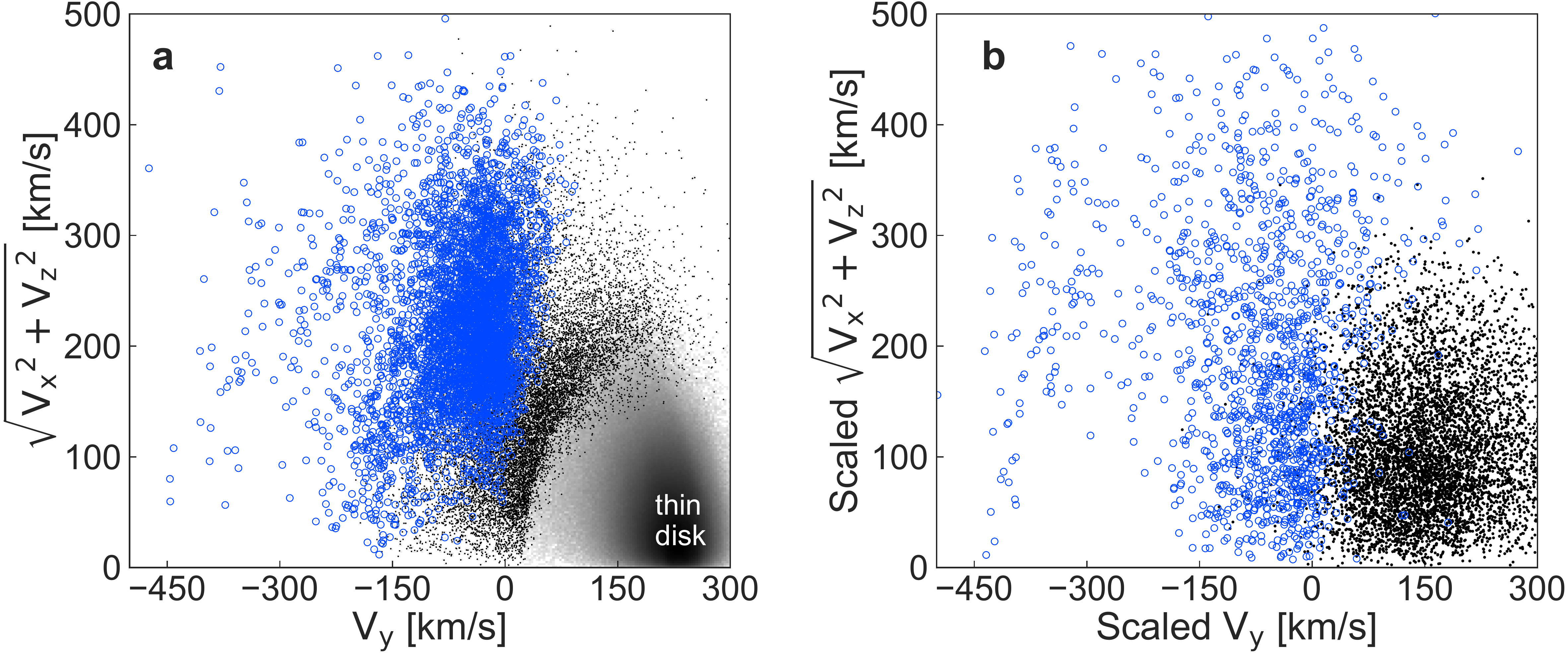}
\caption{\small {\bf Velocity distribution of stars in the Solar
    vicinity in comparison to a merger simulation}. In the left panel,
  the velocities of stars in the disk are plotted with grey density
  contours (because of the large number of stars), while the halo
  stars (selected as those with
  $|{\bf v} - {\bf v}_{LSR}| > 210$~km/s, where ${\rm v}_{LSR}$ is the
  velocity of the Local Standard of Rest) are shown as points. The
  blue points are part of a prominent structure with slightly
  retrograde mean rotational motion, and have been selected here as
  those having $-1500 < L_z < 150$~kpc~km/s and energy
  $E > -1.8 \times 10^5$~km$^2$/s$^2$ (see Methods for details). The
  panel on the right shows the distribution of star particles in a
  small volume extracted from a simulation\cite{villalobos2008} of the
  formation of a thick disk via a 5:1 merger between a satellite (in
  blue) and a pre-existing disk (in black). The overall morphology and
  the presence of an arch (from $V_y \sim -450$~km/s and
  $V_\perp = \sqrt{V_x^2 + V_z^2} \sim 50$~km/s to
  $V_y \sim -150$~km/s and $V_\perp \sim 300$~km/s seen in the left
  panel) can be reproduced qualitatively after appropriately scaling
  the velocities (see Methods), in a simulation where the satellite is
  disky (rather than spherical, as the arch-like feature is sharper),
  and on a retrograde orbit inclined by $\sim 30^{\rm o}$ to
  $60^{\rm o}$.}\label{fig:vel_comp}
\end{figure}

The sharp view provided by the second data release (DR2) of the \gaia
mission\cite{brown2018}, has recently revealed\cite{helmer2018} that,
besides a few tight streams, a significant fraction of the halo stars
near the Sun are associated with a single large kinematic structure
that has slightly retrograde mean motion and which dominates the
Hertzsprung-Russell diagram's (HRD) blue sequence revealed in the
\gaia data\cite{Babusiaux2018}. This large structure is readily
apparent (in blue) in Fig.~\ref{fig:vel_comp}a, which shows the
velocity distribution of stars (presumably belonging to the halo) in
the Solar vicinity inside a volume of 2.5 kpc radius from \gaia data
(see Methods for details). Fig.~\ref{fig:vel_comp}b shows the velocity
distribution from a simulation of the formation of a thick disk via a
20\% mass-ratio merger\cite{villalobos2008}. The similarity between
the panels suggests that the retrograde structure could be largely
made up of stars originating in an external galaxy that merged with
the Milky Way in the past.

\begin{figure}[!ht]
\centering%
\includegraphics[scale=0.45,trim={0.5cm 5cm 0 3cm},clip]{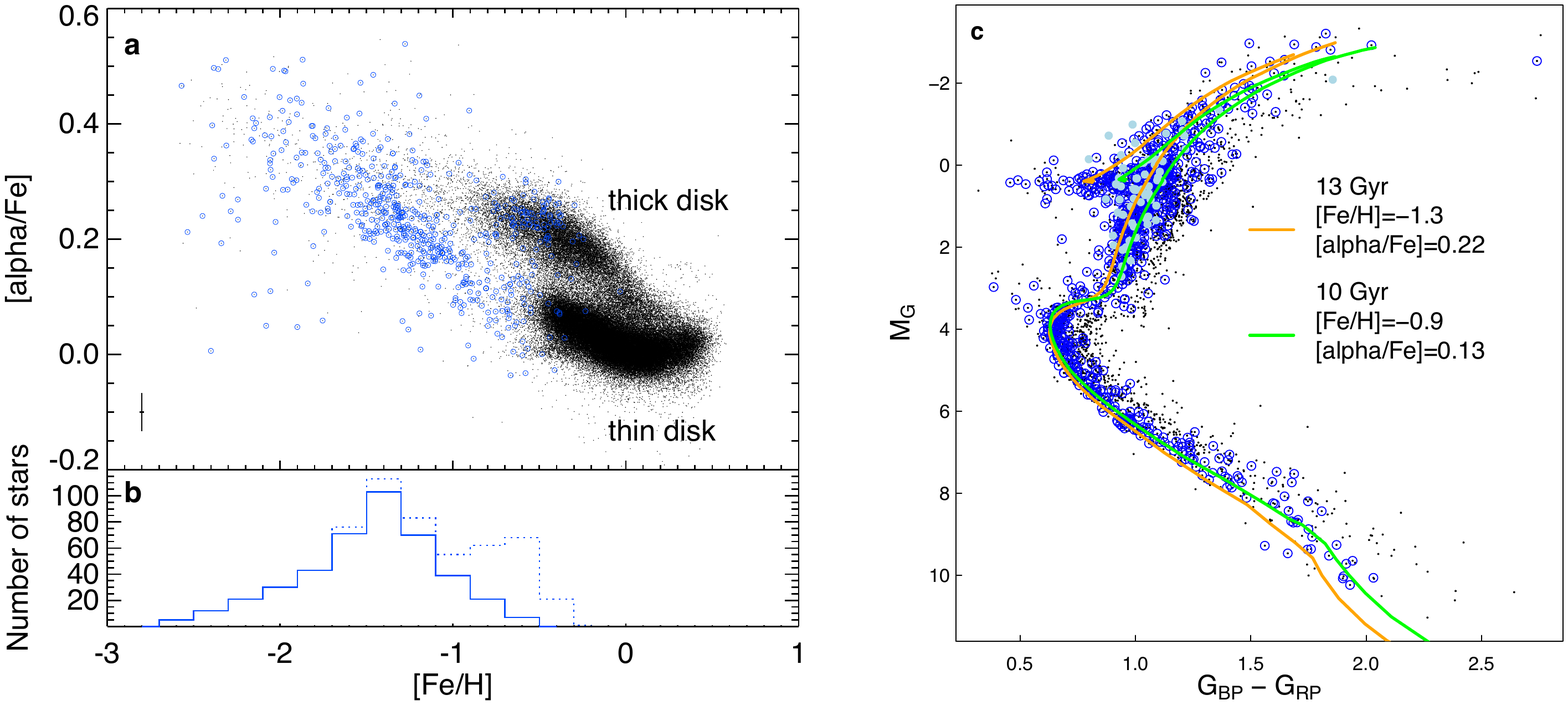}
\caption{\small {\bf Astrophysical properties of stars in
    Gaia-Enceladus.} Panel a) shows the chemical abundances for a
  sample of stars located within 5 kpc from the Sun resulting from the
  cross-match between \gaia and APOGEE. The blue circles correspond to
  590 stars that have $-1500 < L_z < 150$~km/s~kpc and
  $E > -1.8 \times 10^5$~km$^2$/s$^2$ (as in Fig.~\ref{fig:vel_comp}a,
  but now for a larger volume to increase the sample size, see
  Methods). Note the clear separation between the thick disk and the
  sequence defined by the majority of the stars in the retrograde
  structure, except for a small amount of contamination (17\%) by
  thick disk stars (i.e. on the $\alpha$-rich sequence) that share a
  similar phase-space distribution as the structure. The error
    bar in the lower left corner shows the median error for the
    sample. The solid (dotted) histogram in panel b) shows the
  metallicity distribution of the structure without (with) the subset
  of $\alpha$-rich stars. Their distribution peaking at
  [Fe/H]$\sim -1.6$, is very reminiscent of that of the stellar
  halo\cite{helmi2008}. Panel c) is the HRD for halo stars (black
  points, selected as in Fig.~\ref{fig:vel_comp}a with the additional
  photometric quality cuts\cite{Babusiaux2018}: $E(B-V) <0.015$ to
  limit the impact in the magnitudes and colours to less than
  0.05~mag, and {\tt phot-bp-rp-excess-factor}
  $< 1.3+0.06(G_{\rm BP}-G_{\rm RP})^2$) and shows {\it Gaia}'s blue
  and red sequences. Gaia-Enceladus stars are plotted with dark blue
  symbols, with those in APOGEE within 5 kpc and with [$\alpha$/Fe]
  $<-0.14-0.35$ [Fe/H], in light blue. The superimposed
    isochrones\cite{Marigo2017} based on previous
    work\cite{Hawkins2014} show that an age range from 10 to 13 Gyr
    is compatible with the HRD of Gaia-Enceladus.}\label{fig:aFe}
\end{figure}

Support for this hypothesis comes from the chemical abundances of
stars provided by the APOGEE survey\cite{apogee-dr14}. In
Fig.~\ref{fig:aFe}a we plot the [$\alpha$/Fe] vs [Fe/H] abundances for
a sample of stars cross-matched to \gaia DR2 (see Methods for
details).  $\alpha$-elements are produced by massive stars that die
fast as supernovae (SNII), while iron, Fe, is also produced in SNI
explosions of binary stars. Therefore in a galaxy, [$\alpha$/Fe]
decreases with time (as [Fe/H] increases).  Fig.~\ref{fig:aFe}a shows
the well-known sequences defined by the thin and thick disks. The vast
majority of the retrograde structure's stars (in blue), follow a
well-defined separate sequence that extends from low to relatively
high [Fe/H].  (The presence of low-$\alpha$ stars with retrograde
motions in the nearby halo has in fact been reported
before\cite{nissen2010,nissen2011} but for a small sample. The
existence of a well-populated sequence with lower [$\alpha$/Fe] was
demonstrated very recently using also APOGEE data\cite{hayes2018}). An
independent analysis\cite{haywood2018} has confirmed the relation
between {\it Gaia}'s HRD blue sequence and the kinematic structure
shown in Fig.~\ref{fig:vel_comp}a, and established firmly the link to
the low-$\alpha$ stars using both earlier data\cite{nissen2010} as
well as APOGEE, thereby putting the accretion hypothesis on more secure
ground.

The large metallicity spread of the retrograde structure stars 
depicted in Fig.~\ref{fig:aFe}b, implies that they did not form in a
single burst in a low mass system. Furthermore because the more
metal-rich stars have lower [$\alpha$/Fe] at the characteristic
metallicity of the thick disk ([Fe/H]$\sim -0.6$), this means that
they were born in a system with a lower star formation rate than the
thick disk.  The star formation rate required to match the
$\alpha$-poor sequence of the APOGEE data has recently been calculated
using a chemical evolution model and including different elemental
abundances\cite{fernandez2018}, and found to be $0.3 \msol$/yr lasting
for about 2~Gyr. This implies a stellar mass for the progenitor system
of $\sim 6 \times 10^8\msol$, a value that is consistent with the
large fraction of nearby halo stars being associated with the structure
given estimates of the local halo density\cite{helmi2008}, and which
is comparable to the present-day mass of the Small Magellanic
Cloud\cite{vdM}. Interestingly, previous work\cite{hayes2018} has
shown that the trends in the abundances of low metallicity stars in
the Large Magellanic Cloud actually overlap quite well with the
sequence, implying that the structure was comparable to the Large
Magellanic Cloud in its early years.  Furthermore and perhaps even
more importantly, because [$\alpha$/Fe] must decrease as [Fe/H]
increases, the stars in the structure could not have formed in the
same system as the vast majority of stars in the Galactic thick
disk. They must have formed, as previously
suggested\cite{nissen2010,helmer2018,haywood2018}, in a separate
galaxy, which we refer to as Gaia-Enceladus hereafter (see Methods for
the motivation behind the naming).

We now explore whether the Gaia-Enceladus galaxy could have been
responsible at least partly for the formation of the thick
disk\cite{belokurov2018,helmer2018,haywood2018}, as the comparison
between the data and the simulation shown in Fig.~\ref{fig:vel_comp}
would suggest. In that case, a pre-existing disk must have been in
place at the time of the merger. Fig.~\ref{fig:aFe}c plots the HRD of
the halo stars in Fig.~\ref{fig:vel_comp}a showing the Gaia-Enceladus
stars (in blue) populating {\it Gaia}'s blue
sequence\cite{Babusiaux2018,helmer2018,haywood2018}. The thinness
  of this sequence is compatible with an age range from $\sim 10$ to
  13 Gyr given the stars' abundance sequence, as indicated by the
  plotted isochrones\cite{Marigo2017}.  Previous
  studies\cite{schuster2012,Hawkins2014}, on which this age range is
  based, have shown that the stars on the $\alpha$-poor sequence are
  younger than those on the $\alpha$-rich sequence for
  $-1 < $~[Fe/H]~$< -0.5$. This implies that the progenitor of the
Galactic thick disk was in place when Gaia-Enceladus fell in, which
based on the ages of its youngest stars, would suggest that the merger
took place around 10 Gyr ago, i.e. at redshift $z \sim 1.8$.

\begin{figure}[!h]
\centering%
\includegraphics[scale=0.4,trim={0cm 0cm 0cm 0cm},clip]{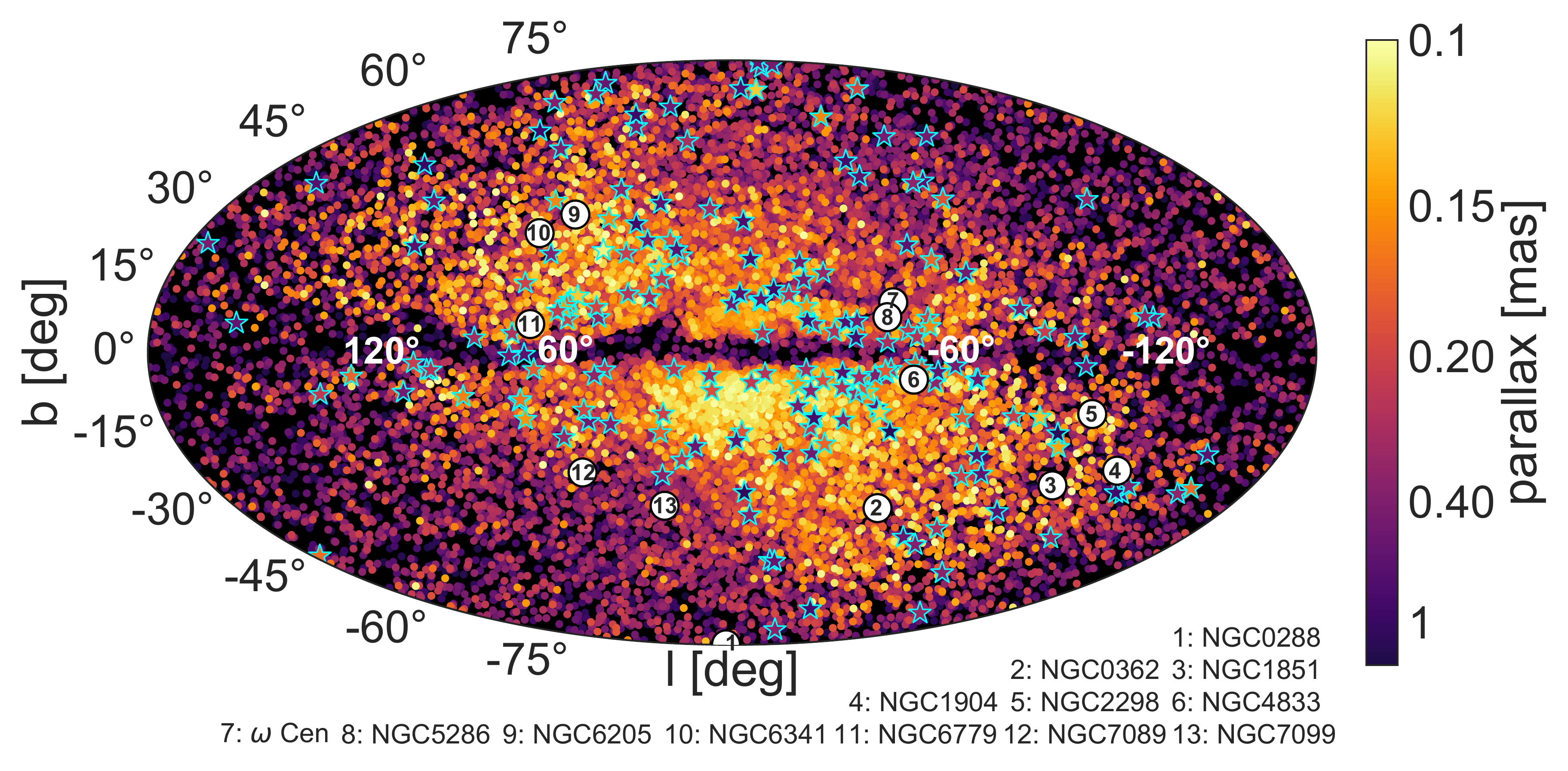}
\caption{\small {\bf Sky distribution of tentative Gaia-Enceladus
    members from a \gaia subsample of stars with full phase-space
    information}. These stars have $\varpi > 0.1$~mas, relative
  parallax error of 20\%, and are colour-coded by their distance from
  the Sun (from near in dark red to far in light yellow). They satisfy
  the condition $-1500 < L_z < 150$~kpc~km/s.  Because of the larger
  volume explored, we do not include additional selection criteria
  based on energy, as done for Fig.~\ref{fig:aFe} (since energy
  depends on the Galactic potential whose spatial variation across the
  volume explored is less well-constrained than its local value), nor
  on velocity as for Fig.~\ref{fig:vel_comp}a (because of spatial
  gradients). We thus expect some amount of contamination by thick
  disk stars, especially towards the inner Galaxy (see Methods). The
  starry symbols are \gaia RR Lyrae stars potentially
  associated to this structure. To identify these, we bin the sky in
  128$\times$128 elements, and $\log \varpi$ in bins of 0.2 width
  (mimicking the relative parallax error), and measure the average
  proper motion of Gaia-Enceladus stars in each 3D bin. We then
  require that the RR Lyrae have the same proper motion (within 25
  km/s in each component at their distance), which for example
  corresponds to 1 mas/yr for those with $\varpi \sim 0.2$~mas.
  Globular clusters with $L_z < 250$~kpc~km/s, located between 5
    and 15 kpc from the Sun, and 40$^{\rm o}$ away from the
    Galactic centre, are indicated with solid
  circles.}\label{fig:sky-all}
\end{figure}
\begin{figure}[!h]
\centering%
\includegraphics[scale=0.28,trim={0cm 0cm 0cm 0cm},clip]{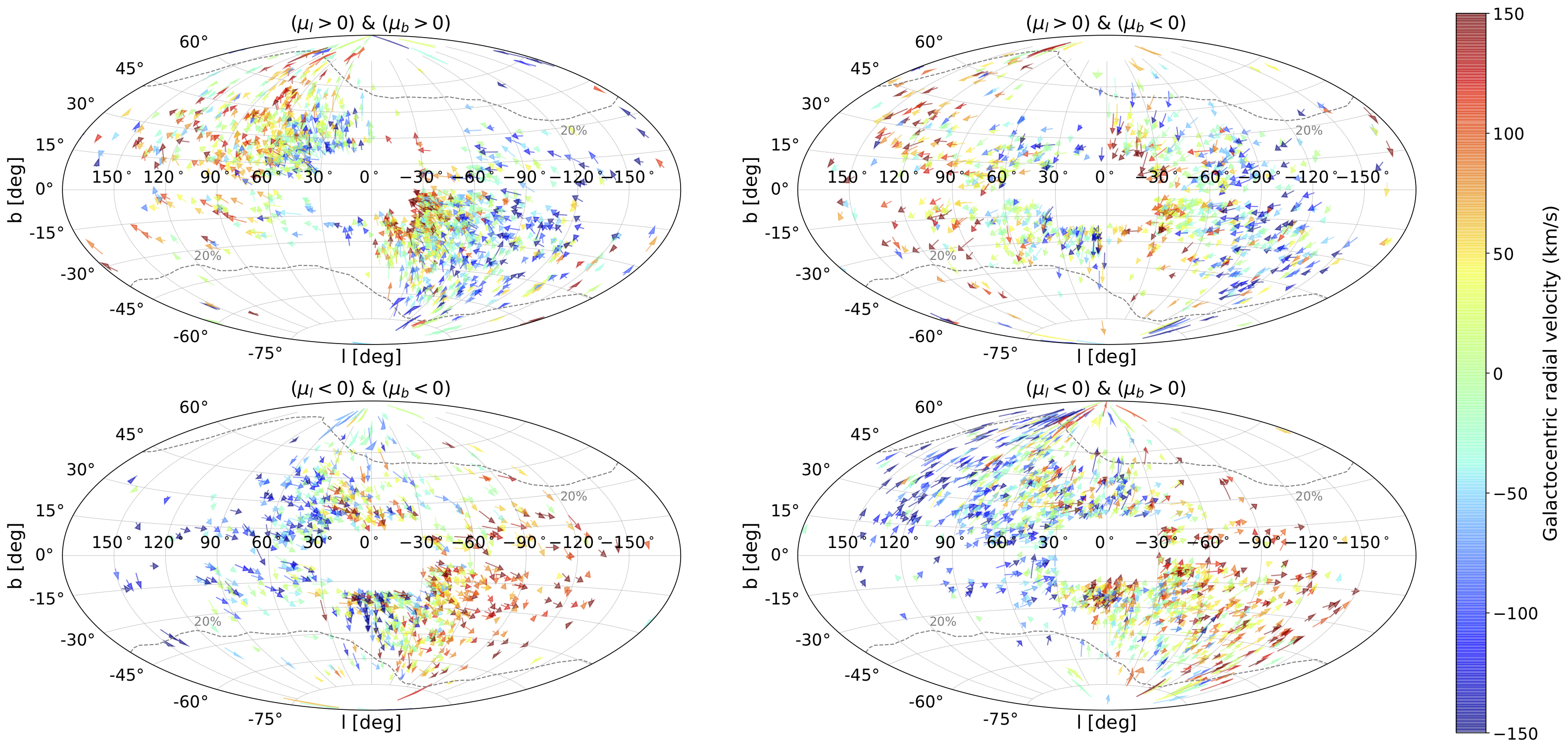}
\caption{\small {\bf Kinematic properties of Gaia-Enceladus tentative
    members on the sky.}  The plotted stars are a subset of those in
  Fig.~\ref{fig:sky-all} with $0.1 < \varpi < 0.2$~mas, and are colour
  coded by their radial velocity with the arrows indicating their
  direction of motion.  To avoid cluttering the panels correspond to
  different proper motion ranges, and we have removed stars close to
  the bulge (within 30$^{\rm o}$ in longitude and 20$^{\rm o}$ in
  latitude). All velocities have been corrected for the Solar and for
  the Local Standard of Rest motions. The grey contours encompass
    90\% of the stars in the 6D \gaia set with
    $0.1 < \varpi < 0.2$~mas and having 20\% relative parallax error,
    and clearly demonstrate this selection criterion impacts our
    ability to identify distant Gaia-Enceladus stars in certain
    regions of the sky. Notice the large-scale pattern in the radial
  velocity, as well as its correlation with the proper motion
  component $\mu_l$: stars with $\mu_l > 0$ (top panels) have
  $v_{GSR} > 0$ for $l \gtrsim 75^{\rm o}$ and $v_{GSR} < 0$ for
  $l \lesssim -75^{\rm o}$, while the opposite occurs for $\mu_l <
  0$. Such a global pattern (and its reversal for
  $-75^{\rm o} \lesssim l \lesssim 75^{\rm o}$), arises because of the
  coherent retrograde sense of rotation of the stars in their orbits
  (i.e. they have mostly $L_z \lesssim 0$), but the correlation with
  $\mu_l$ is a result of their elongated orbits, e.g. we see that if
  $\mu_l > 0$ and $l \gtrsim 75^{\rm o}$ stars are typically moving
  outwards with high speed and away from the Solar radius
  ($v_{GSR} \gtrsim 100$ km/s).}
\label{fig:sky-pm}
\end{figure}

 Such a prominent merger must have left debris over a large volume of
the Galaxy. To explore where we may find other tentative members of
Gaia-Enceladus beyond the solar neighbourhood, we consider stars in
the \gaia 6D sample with 20\% relative parallax error, with
$\varpi > 0.1$~mas and having $-1500 < L_z < 150$~kpc~km/s.
Fig.~\ref{fig:sky-all} shows that nearby tentative Gaia-Enceladus
stars (with $\varpi > 0.25$~mas, darker points) are distributed over
the whole sky, this subset being more than 90\% complete. More distant
stars are preferentially found in specific regions of the sky, and
although for such small \gaia parallaxes ($\varpi = 0.1 - 0.25$~mas)
the zero-point offset ($\sim -0.03$~mas) is significant and this
affects the selection in $L_z$, it does not to the extent that it can
produce the observed asymmetry on the sky. At least in part this
asymmetry is due to the 20\% relative parallax error cut, as
highlighted in Fig.~\ref{fig:sky-pm} (see Methods for more details and
also for possible links to known overdensities). In Figure
\ref{fig:sky-all} we have also overplotted (with starry symbols) a
subset of 200 \gaia RR Lyrae stars\cite{variables2018}. These have
proper motions similar to the mean of the candidate Gaia-Enceladus
stars with full phase-space information, at their sky position and
parallax.  Thirteen globular clusters can also be associated to
Gaia-Enceladus on the basis of their angular momenta\cite{helmi2018}
(Fig.~\ref{fig:sky-all}). All these clusters show a consistent
age-metallicity relation\cite{gc-age-met}.

Fig.~\ref{fig:sky-pm} shows the velocity field of the more distant
stars associated to Gaia-Enceladus. Notice the large-scale gradient in the
radial velocity across the full sky. Such a coherent pattern can only
be obtained if stars are moving in the same (retrograde) direction on elongated
orbits. The proper motions, depicted by the arrows, reveal a rather
complex velocity field. This is expected, given the large mass
of the progenitor object and the short mixing timescales in the inner
Galaxy\cite{hws2003}. Nonetheless, in this complexity we see streams:
close stars often move in the same direction.  This is a very
significant effect as established by comparing to mock sets
constructed assuming a multivariate Gaussian for the velocities (see
Methods for details).

We conclude that the halo near the Sun is strongly dominated by a
single structure of accreted origin, as hinted also by other
work\cite{belokurov2018,helmer2018}, and leaving little room for an
in-situ contribution\cite{haywood2018}. It is however, not necessarily
representative of the whole stellar halo, as debris from other
accreted large objects (with e.g. different chemical abundance
patterns) might dominate elsewhere in the Galaxy.  We also
conclude that the Milky Way disk experienced a significant merger in
its history. We estimate the mass-ratio of this merger at the time it
took place as
$\frac{\displaystyle M_{\rm vir}^{GE}}{\displaystyle M_{\rm vir}^{MW}}
= \frac{\displaystyle f^{MW}}{\displaystyle f^{GE}} \times
\frac{\displaystyle M_*^{GE}}{\displaystyle M_*^{MW}}$, where $f$ is
the ratio of the luminous-to-halo mass of the object. At the
present-time, $f^{MW,0} \sim 0.04$ for the Milky
Way\cite{mcmillan2017}, and if we assume that Gaia-Enceladus would be
similar to the Large Magellanic Cloud had it evolved in isolation,
then $f^{GE,0} \sim 0.01$\cite{vdM}. It has been shown\cite{behroozi}
that the redshift evolution of $f$ between $z=2$ and $z=0$ for objects
of the Magellanic Cloud scale and the Milky Way is similar, implying
that $f^{MW}/f^{GE} = f^{MW,0}/f^{GE,0} \sim 4$. Therefore, taking
$M_*^{MW}$ at the time of the merger to be the mass of the thick
disk\cite{mcmillan2017}, i.e. $\sim 10^{10}\msol$, we obtain a
mass-ratio for the merger of $\sim 0.24$. This implies that the
merging of Gaia-Enceladus must have led to significant heating and to
the formation of a thick(er) disk.

\bibliographystyle{unsrtnat}

{\bf Acknowledgements.} We are grateful to {\' A}. Villalobos for
permission to use his suite of simulations, and to M. Breddels for the
software package Vaex ({\tt
  http://vaex.astro.rug.nl/}), partly used for our
analyses. We thank H.-W.~Rix, D.~Hogg and A.~Price-Whelan for
comments. We have made use of data from the European Space Agency
mission \gaia \linebreak ({\tt http://www.cosmos.esa.int/gaia}), processed by the
\gaia Data Processing and Analysis Consortium (DPAC, see 
{\tt http://www.cosmos.esa.int/web/gaia/dpac/consortium}). Funding for
DPAC has been provided by national institutions, in particular the
institutions participating in the \gaia Multilateral Agreement. We
have also made use of data from the APOGEE survey, which is part of
Sloan Digital Sky Survey IV. SDSS-IV is managed by the Astrophysical
Research Consortium for the Participating Institutions of the SDSS
Collaboration ({\tt http://www.sdss.org}).  AH acknowledges financial
support from a Vici grant from the Netherlands Organisation for
Scientific Research (NWO), and AB from the Netherlands Research School
for Astronomy (NOVA).

{\bf Authors contributions.} All the authors critically contributed to
the work presented here.  AH led and played a part in all aspects of
the analysis, and wrote the manuscript.  CB compiled the APOGEE data,
provided the cross-match to the \gaia data, was instrumental for the
chemical abundance aspects, and together with DM analysed the
Hertzsprung-Russell diagram. HHK and JV carried out the dynamical
analysis and identification of member stars. AB triggered this paper,
explored the impact of selection effects, and contributed to its
writing together with the other co-authors.

The authors declare that they have no competing financial
interests. Reprints and permissions information are available at
{\tt www.nature.com/reprints}. Correspondence and requests for materials
should be addressed to A. Helmi: ahelmiATastro.rug.nl.

\appendix

\begin{center}
{\bf Methods}
\end{center}

We describe here the motivation behind the name Gaia-Enceladus.  In
Greek mythology Enceladus was one of the Giants (Titans), and the
offspring of Gaia (which represents the Earth), and Uranus
(representing the Sky). Enceladus was said to be buried under Mount
Etna and responsible for earthquakes in the region. The analogies to
the accreted galaxy reported and characterized in this paper are many,
and they include: i) being offspring of Gaia and the sky, ii) having
been a ``giant'' compared to other past and present satellite galaxies
of the Milky Way, iii) being buried (in reality first disrupted by the
Milky Way and then buried, also in the \gaia data as it were), and iv)
being responsible for seismic activity (i.e. shaking the Milky Way and
thereby leading to the formation of its thick disk). We refer to the
accreted galaxy as Gaia-Enceladus to avoid confusion with one of
Saturn's moons, also named Enceladus.

\section{Dataset, selection criteria and the effect of systematics}
\label{sec:data}

For the work presented in the main section of the paper, we selected
stars from the \gaia 6D-dataset\cite{brown2018} with small relative
parallax error $\varpi/\sigma_\varpi > 5$, which allows us to compute
their distance as $d = 1/\varpi$. For Figure~\ref{fig:vel_comp}, we consider only stars
with $\varpi > 0.4$~mas (i.e. within 2.5~kpc from the Sun) to limit
the impact of velocity gradients. The velocities were obtained using
the appropriate matrix transformations form the observables
$\alpha, \delta$, $\mu_{\alpha*}$, $\mu_\delta$, $v_{los}$ and
distances $d$. These velocities have then been corrected for the
peculiar motion of the Sun\cite{schoenrich} and the Local Standard of
Rest velocity, assuming a value\cite{mcmillan2017} of
$V_{LSR} = 232$~km/s.

We select halo stars (such as the black points in
Figure~\ref{fig:vel_comp}a) as those that satisfy
$|{\bf v} - {\bf v}_{LSR}| > 210$~km/s\cite{helmer2018}. This
condition is an attempt to remove the contribution of the disk(s),
although towards the inner Galaxy, this is less effective because of
the increasing velocity dispersion of disk stars\cite{Katz2018}.  To
select members of the retrograde structure (such as the blue points in
Figure~\ref{fig:vel_comp}a), 
we inspect the energy vs $L_z$ distribution of the stars in our
dataset. The energy is computed assuming a Galactic potential
including a thin disk, bulge and halo components\cite{Helmi2017}. For
example, the left panel of Extended Data Fig.~1 shows the energy vs
$L_z$ distribution for all halo stars within 5~kpc from the Sun
($\varpi > 0.2$~mas). We have here removed stars with {\tt
  phot-bp-rp-excess-factor} $ >1.27$ (this is enough to remove some
not so well-behaved globular cluster stars so we do not apply a
colour-dependent correction\cite{Arenou2018}). This figure shows that
the regions occupied by the retrograde structure and by the disk are
relatively well-separated. There is however some amount of overlap,
particularly for higher binding energies and lower angular
momenta. Therefore even the selection criteria of $L_z < 150$~kpc~km/s
and $E > -1.8 \times 10^5$~km$^2$/s$^2$, indicated by the
straightlines, will not yield a pure (thick disk free) sample of stars
in the structure. This figure reveals also the large range of energies
in the structure, indicating that member stars are expected over a
large range of distances.

Because the energies of stars depend on the gravitational potential of
the Galaxy, and its form and amplitude are not so well-constrained
beyond the Solar neighbourhood, we use a criterion based only on $L_z$
to find additional members of the structure/Gaia-Enceladus beyond the
immediate vicinity of the Sun (as for Fig.~\ref{fig:sky-all} of the main
section).  The central panel of Extended Data Fig.~1 shows the $L_z$
vs Galactocentric distance in the disk plane $R$, for all stars in the
\gaia 6D-dataset with $\varpi/\sigma_\varpi > 5$, and 
including stars with parallaxes $\varpi > 0.2$~mas.  This plot shows
that a selection based only on $L_z$ works relatively well to isolate
Gaia-Enceladus stars near the Sun and also farther out in the Galaxy. However
for the inner regions there is much more overlap and hence the
distinction between the thick disk and Gaia-Enceladus is less
straightforward, and the amount of contamination by thick disk stars
is likely to be much higher. Furthermore, we expect the orbits of some
stars in the progenitor of the thick disk to have been perturbed so
significantly during the merger\cite{baptiste} that they will
``mingle'' with those from Gaia-Enceladus.

The rightmost panel of Extended Data Fig.~1 shows the $z$-angular
momentum as function of cylindrical radius of stellar particles in a
simulation of the merger of a pre-existing disk and a massive
satellite\cite{villalobos2008,villalobos2009} (the same of
Fig.~\ref{fig:vel_comp}b). The example here corresponds to the
redshift $z=1$ simulation of a disk with
$M_* = 1.2 \times 10^{10} \msol$ and a satellite with
$M_{*,sat} = 2.4 \times 10^{9} \msol$. Because of the lower host mass
used in this simulation (compared to the present-day mass of the Milky
Way), the spatial scales and velocities are typically smaller compared
to the data. Therefore in the simulations, we consider as solar
vicinity a volume centered at
$R_{\rm sun}^{sim} = 2.4 \times R^{final}_{\rm thick}$, where
$R^{final}_{\rm thick} = 2.26$~kpc\cite{villalobos2008}. We also scale
the positions by $R_{\rm sun}/R_{\rm sun}^{sim} = 1.5$ and the
velocities by $v_{\rm thick,sun} /v^{final,sim}_{\rm thick}$, where
$v_{\rm thick,sun} = 173$~km/s is the rotational velocity of the thick
disk near the Sun\cite{heather1990} and $v^{final,sim}_{\rm thick}$ is
that of the thick disk in the simulation at $R_{\rm
  sun}^{sim}$. Extended Data Fig.~1c shows that like for the data,
the separation between accreted and host disk stars is less effective
for small radii.

For Fig.~\ref{fig:aFe}a, we have cross-matched the catalogues from
\gaia DR2 and APOGEE\cite{apogee-dr14,apogee-majewski} DR14 and
retained only stars with estimated distances from both these
catalogues (i.e. spectrophotometric and trigonometric parallaxes)
consistent with each other at the 2$\sigma$ level. We also impose a
relative parallax error of 20\%.  More than 100,000 stars within 5 kpc
from the Sun satisfy these conditions. The abundances shown in
Fig.~\ref{fig:aFe} stem from the ASCAP pipeline\cite{ASCAP}.

The presence of a parallax zero-point offset in the \gaia
data\cite{Lindegren2018} has been established thoroughly, and is
partly (if not only) due to a degeneracy between the parallax and the
basic-angle variation of the \gaia satellite\cite{bam}. Its amplitude
varies with location on the sky\cite{Arenou2018,Lindegren2018}, and is
on average $-0.029$~mas and has an RMS of
$\sim 0.03$~mas\cite{helmi2018}. Such variations make it very
difficult to perform a correction a posteriori for the full \gaia DR2
dataset (although the expectation is that its effect will be smaller
for \gaia DR3). The discovery and characterization of Gaia-Enceladus
was done using stars with parallaxes $\varpi > 0.4$~mas for
Fig.~\ref{fig:vel_comp} of the main section, and in Fig.~\ref{fig:aFe}
for stars with $\varpi > 0.2$~mas from the cross-match of \gaia and
APOGEE. We therefore expect the derived
kinematic and dynamical quantities for these subsets to be largely
unaffected by the systematic parallax error. However, for
Figs.~\ref{fig:sky-all} and \ref{fig:sky-pm} of the main section
of the paper, we selected stars on the basis of their $L_z$ although
we focused on properties which are independent of the parallax, such
as position on the sky and proper motions. Nonetheless, to establish
how important the parallax zero-point offset is on the selection via
$L_z$ we perform the following test.

We use the \gaia Universe Model Snapshot GUMS v18.0.0\cite{gums}, and
select stars according to the following criteria: $6 \le G \le 13.0$,
$0.2 \le \log g \le 5$ and $3000 \le T_{eff} \le 9000$~K. This
selection leads to a total of 7403454 stars distributed across all
Galactic components. For these stars we compute error-free velocities
and $L_z$. We convolve their true parallax with a Gaussian with a
dispersion of depending on the magnitude of the
star\cite{http_gaia}. The parallax is reconvolved with a Gaussian with
a mean of $-0.029$~mas and a dispersion of
0.030~mas. Using these observed
parallaxes, we compute ``observed'' velocities and $L_z$.

We find that for measured distances smaller than 5 kpc, there is no
shift in the derived $L_z$, while for a shell between 5 and 7.5 kpc
the median amplitude of the shift is $\sim -50$~kpc~km/s, making the
observed $L_z$ more retrograde. For a shell between 9 and 10 kpc, the
median shift is small and has an amplitude of $20$~kpc km/s,
presumably reflecting that at such large distances, the random errors
on the individual stars' measurements dominate. The results are shown
in the left panel of Extended Data Fig.~2 where we plot the difference
between the true (initial) and ``measured'' distributions of $L_z$ for
stars ``observed'' to be located at distances between 5 and 10 kpc,
for $l = (-60^{\rm o}, -20^{\rm o})$. The panel on the right shows the
distribution of the mean value of the difference over the whole sky,
and although it reveals certain patterns, these are different from
those seen in Fig.~\ref{fig:sky-all}.  As stated in the main
  section of the paper, the lack of distant stars in the regions
  outside of the contours plotted in Fig.~\ref{fig:sky-pm}, is the
  result of a quality cut in the relative parallax error of 20\%.
  This selection criterion allows for parallax errors in the
  range 0.02 to 0.04 mas for the most distant stars (with
  $0.1 < \varpi < 0.2$~mas), and these are only reached in those regions
  of the sky that have been surveyed more frequently by {\it Gaia},
  such as around the ecliptic poles. The \gaia RR Lyrae stars
  associated to Gaia-Enceladus suffer of course also from this effect,
  as a lower number of visits leads to more difficult identification
  and hence to lower levels of completeness\cite{variables2018}.

\section{Random sets and significance of features}

To understand how different the dynamical properties of the \gaia 6D
dataset are in comparison to a smooth distribution, we plot the
distribution of velocities in Extended Data Fig.~3a and of $E$ vs
$L_z$ in Extended Data Fig.~3b for randomized datasets. These smooth
datasets have been obtained from the \gaia data shown in
Fig.~\ref{fig:vel_comp}a and in Extended Data Fig.~1a, respectively,
by re-shuffling the velocities. That is, for each star, we assign
randomly a $v_y$ and $v_z$ velocity from two other stars in the
sample. This results in distributions with the same 1D velocity
distributions as the original data, but without any correlations or
lumpiness. The comparison of Fig.~\ref{fig:vel_comp}a to 
Extended Data Fig.~3a shows that the distribution in the random
dataset is indeed much smoother than the data, and that the overall
velocity dispersion in the $y$-direction has increased because there
no longer is a clear separation between the region occupied by
Gaia-Enceladus and by the thick disk. The comparison of
Figs.~\ref{fig:vel_comp}b and Extended Data Fig.~3b is even more
revealing and clearly shows that the structure defined in $E$ vs $L_z$ by
Gaia-Enceladus stars has effectively disappeared in the randomized
dataset. Similar conclusions are reached when, instead of using a
reshuffled dataset, we compare the distributions to those in the GUMS
model.

Fig.~4 of the main section of the paper shows the radial velocities
and proper motions (corrected for the Solar and for the Local Standard
of Rest motions) for stars with $0.1 < \varpi < 0.2$~mas and
$-1500 < L_z < 150$~kpc~km/s. These stars are tentative members of
Gaia-Enceladus, although as discussed earlier towards the inner Galaxy
contamination by thick disk stars becomes more important for large
distances. The arrows depicting the proper motions suggest that stars
that are closeby on the sky move in similar directions. We establish here whether
this is significant by comparing to a mock dataset.

The mock dataset uses the measured positions of the stars that are
plotted in Fig.~4, but their velocities are generated randomly
according to a multivariate Gaussian distribution with dispersions in
$v_R$, $v_\phi$ and $v_z$ of 141, 78 and 94 km/s
respectively\cite{posti2018}.  During the process of generation, we
only keep stars' velocities that satisfy $-1500 < L_z < 150$~kpc~km/s,
as in the real data. To quantify the degree of coherence in the proper
motions of neighbouring stars on the sky, we perform the following
test. For each star, we find its nearest neighbour on the sky, and
then determine the angle between their proper motion vectors. We then
count the number of such pairs having a given
angle. Extended Data Fig.~4 shows the distribution of these pairs
for the \gaia subsample (in blue) and for the mock (in red). There is
a clear excess of pairs of stars with similar directions of motion in
the data in comparison to the mock.

\section{Context and link to other substructures}

Hints of the presence of a population like Gaia-Enceladus have been
reported in the literature in the last two decades, and were typically
based on small samples of stars. These hints were of chemo-dynamical
nature\cite{Chiba-Beers2000,Beers2017,carollo2013} and sometimes
attributed to accretion\cite{morrison2009,brook2003}, but also based
purely on chemical signatures, such as the $\alpha$-poor
sequence\cite{nissen2010,nissen2011}. More recently, cross-matches to
the first data release of the \gaia mission\cite{brown2016} also
revealed the contrast between the metal-rich population supported by
prograde rotation and associated to the tail of the thick
disk\cite{bonaca2017}, and the metal-poor halo, i.e. what we have just
identified as Gaia-Enceladus. Furthermore, in one
study\cite{belokurov2018} the difference in the kinematics of these
two populations, and the measurement of a very radially biased
velocity ellipsoid for halo stars with [Fe/H]~$ > -1.7$, led to the
proposal that this population (which was termed ``Gaia sausage'') could
be the result of a significant merger. Although this could be also
attributed to an in-situ formation via a radial collapse, this
hypothesis gained further supported by their orbits leading to the
break in the halo density profile at $\sim
20$~kpc\cite{Deason2018}. All of these pieces together outline the
case for the discovery and detailed characterization of Gaia-Enceladus
reported here.

The more distant Gaia-Enceladus debris occupies large portions of the
sky not extensively covered by other existing surveys. There is
however, a recent detection of an overdensity identified in PanSTARRS
and WISE with the help of \gaia proper motions\cite{Conroy}, which
overlaps with the northern part of the more distant Gaia-Enceladus
stars for $-2 < \mu_{\alpha} < -1$~mas/yr and $-1 < \mu_{\delta} < 0$~mas/yr, 
and partly (but not fully because of the PanSTARRS footprint)
with the southern part, for $0 < \mu_{\alpha} < 1$~mas/yr and
$-3 < \mu_{\delta} < -1$~mas/yr. There could
potentially be also a relation to the Hercules Aquila Cloud\cite{hac}
identified in SDSS, although this appears to be offset both in the
northern and southern hemispheres and located at a larger
distance. The location on the sky of intermediate distance
Gaia-Enceladus stars would seem to overlap with the Hercules thick
disk cloud\cite{Hercules-Thick-Disk}, especially in the fourth
Galactic quadrant below the Galactic plane.

\bibliographystyle{unsrtnat}

\vspace*{0.5cm}

{\bf Data Availability Statement} All data generated or analysed
during this study are included in this published article as 
Source Data files. 

\vspace*{0.5cm}

\section*{Extended Data Figure Captions}

{\bf Extended Data Figure 1. Slices of phase-space used to isolate
    Gaia-Enceladus stars}. Panel a): Energy $E$ vs $L_z$ for stars in
  the 6D \gaia dataset, satisfying the quality criteria described in
  the text, with $\varpi > 0.2$~mas (5~kpc from the Sun) and with
  $|{\bf v} - {\bf v}_{LSR}| > 210$~km/s.  The straightlines indicate
  the criteria used to select Gaia-Enceladus stars, namely
  $-1500 < L_z < 150$~kpc~km/s and
  $E > -1.8 \times 10^5$~km$^2$/s$^2$. These criteria follow roughly
  the structure's shape (see for comparison Extended Data Fig.~3b),
  but are slightly conservative for the upper limit of $L_z$ to
  prevent too much contamination by the thick disk. However, small
  shifts such as considering an
  upper limit of 250~kpc~km/s or a lower limit of $-750$~kpc~km/s for $L_z$, or
  $E > -2 \times 10^5$~km$^2$/s$^2$ do not result in drastic changes
  to the results presented in the paper. Panel b): $L_z$ vs
  Galactocentric distance $R$ for all stars in the 6D \gaia with
  $\varpi > 0.2$~mas. The black points are the halo sample shown in
  panel a). Panel c): same as panel b) for star particles in the
  merger simulation\cite{villalobos2008} shown in
  Fig.~\ref{fig:vel_comp}b, where blue correspond to the stars from
  the satellite, and grey to the host disk, and the positions and
  velocities have been scaled as described in the text. In this
  figure, the energy has been scaled by $E_{\rm sun}$ (which is
  $-1.63\times10^5$~km$^2$/s$^2$ in the Galactic potential used),
  $L_z$ by $L_{z,{\rm sun}} = 1902.4$~kpc~km/s, and $R$ by the solar
  distance $R_{\rm sun} = 8.2$~kpc.

{\bf Extended Data Figure 2.  Effect of a zero-point offset in the parallax on
    $L_z$}. Panel a) shows the distribution of the difference between
  the initial and ``measured'' (after error convolution) $L_z$ for
  GUMS stars with ``measured'' distances between 5 and 10~kpc and with
  $l = (-60^{\rm o}, -20^{\rm o})$. Panel b) shows the mean value of
  the difference over the full sky. 

{\bf Extended Data Figure 3. Distribution of stars' dynamical properties for a
    smooth dataset}. Panel a) shows the velocity distribution and
  panel b) the $E$ vs $L_z$ distribution for a dataset obtained by
  reshuffling the velocities of the stars plotted in Fig.~\ref{fig:vel_comp}a
  and in Extended Data Fig.~1a, respectively. The visual comparison to those
  figures shows that these random sets are less clumped than the
  observed distributions of the \gaia halo~stars.

  {\bf Extended Data Figure 4. Distribution of angles between the
    proper motion vectors for neighbouring stars on the sky.} The blue
  and red histograms correspond respectively, to 
  Gaia-Enceladus and to a mock dataset. This mock dataset uses the
  positions of the stars in Gaia-Enceladus, but velocities generated
  randomly according to a multivariate Gaussian
  distribution\cite{posti2018}, after which only stars' velocities
  that satisfy $-1500 < L_z < 150$~kpc~km/s are kept, as in the real
  data. For each star, we find its nearest neighbour on the sky, and
  then determine the angle~$\Delta \theta$ between their proper motion
  vectors for the data and for the mock. We then count the number of
  such pairs having a given angle $\Delta \theta$.

\vspace*{2cm}
\section*{Extended Data Figures}
\begin{figure}
\centering%
\includegraphics[scale=0.2,trim={0cm 0cm 0cm 0cm},clip]{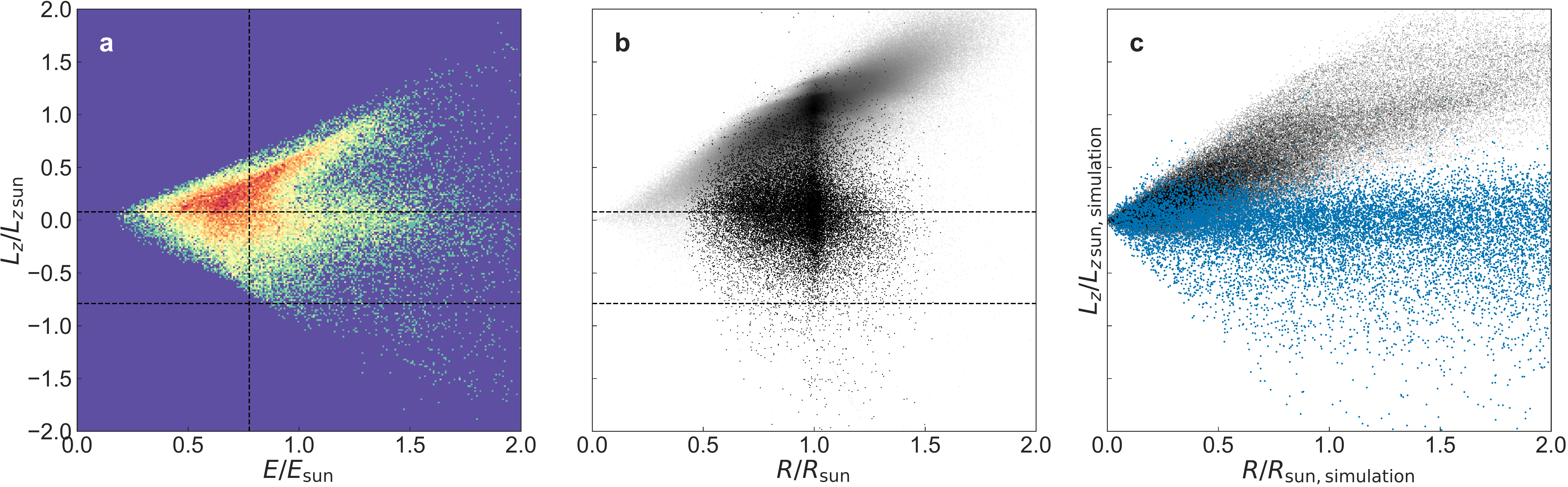}
\mbox{Extended Data Fig. 1}
\end{figure}
\begin{figure}
\centering%
\includegraphics[scale=0.3,trim={0cm 0cm 0cm 0cm},clip]{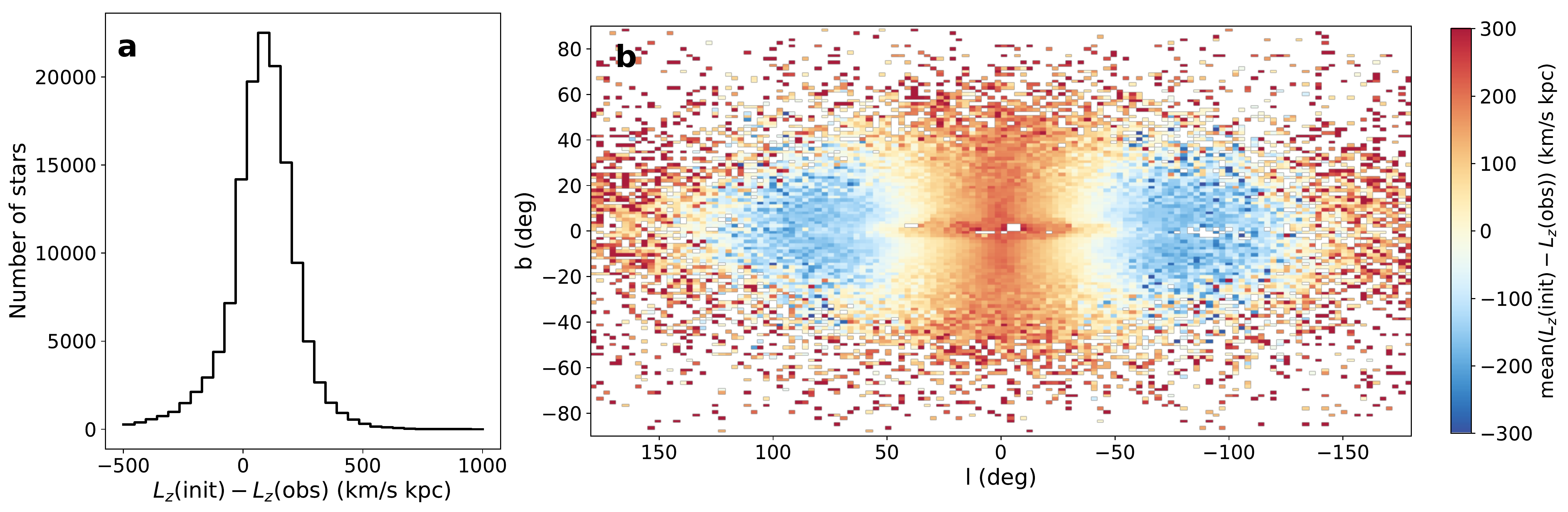}
\mbox{Extended Data Fig. 2}
\end{figure}

\begin{figure}
\centering%
\includegraphics[scale=0.5,trim={0cm 5cm 0cm 2cm},clip]{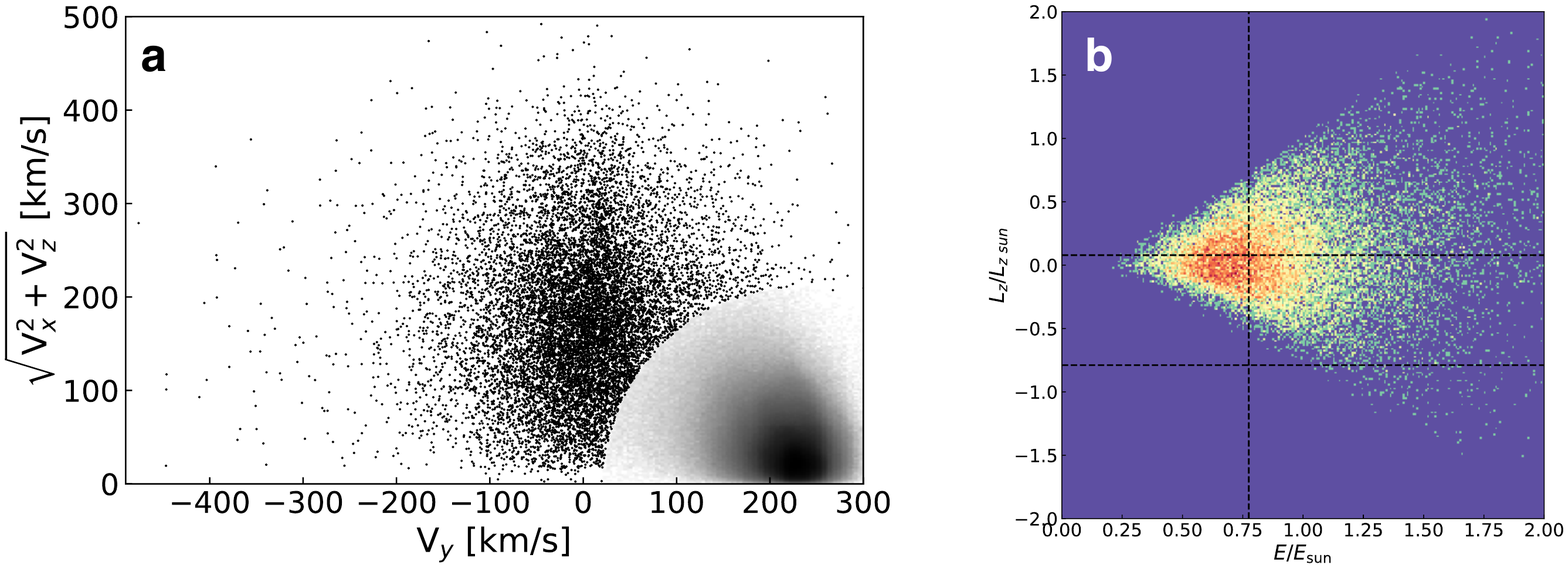}
\mbox{Extended Data Fig. 3}
\end{figure}

\begin{figure}
\centering%
\includegraphics[scale=0.4,trim={0cm 0cm 0cm 0cm},clip]{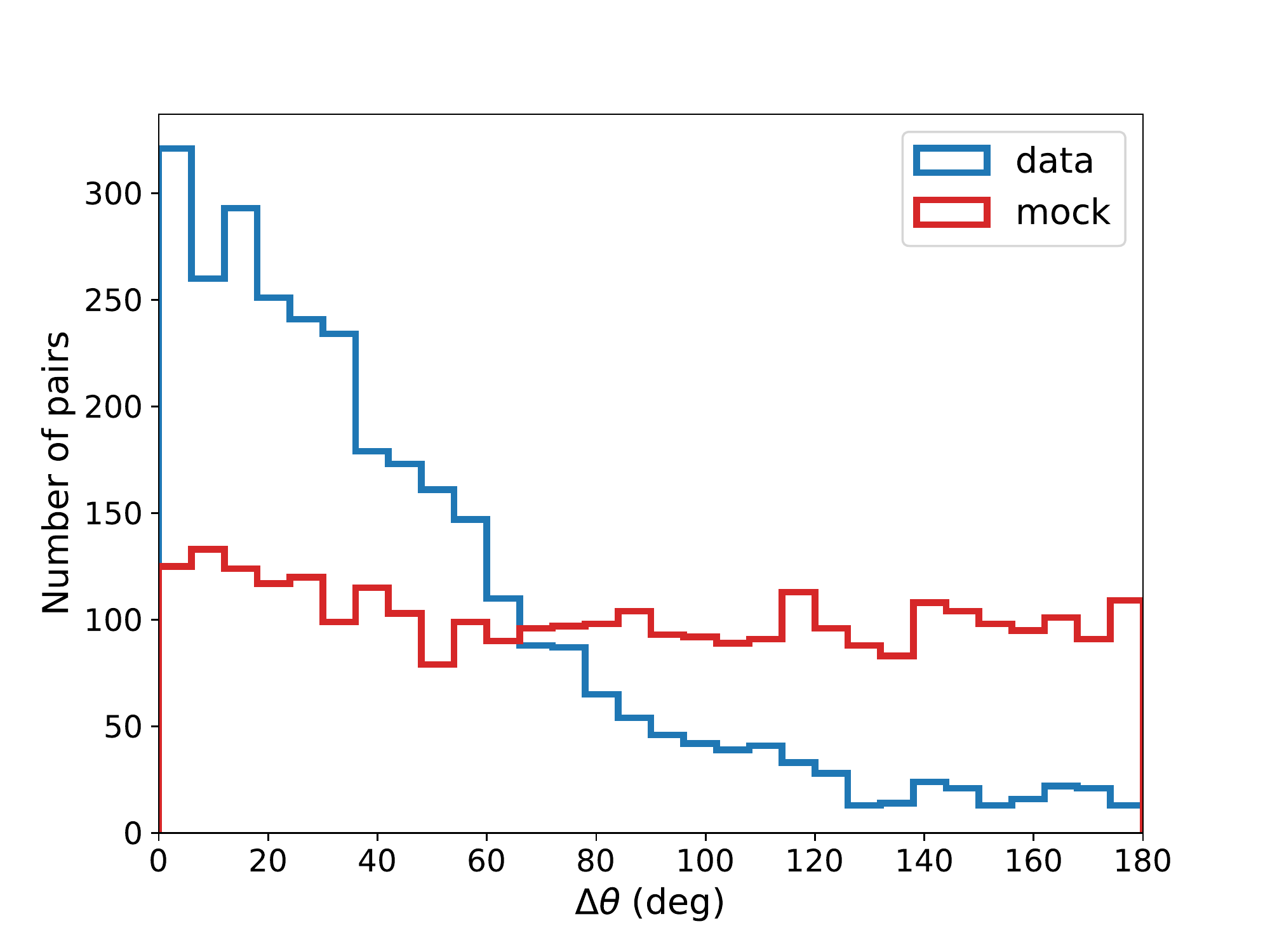}\\
\mbox{Extended Data Fig. 4}
\end{figure}

\end{document}